\documentclass[ preprintnumbers]{revtex4}
\usepackage{graphicx}
\usepackage{epsfig}
\usepackage{bm}
\usepackage{amsmath}
\usepackage{amssymb}
\usepackage{epstopdf}
\begin{document}

\title{Looking for bimodal distributions in multi-fragmentation reactions }

\author{F. Gulminelli}
\affiliation{~LPC (IN2P3-CNRS/Ensicaen et Universit\'e), F-14076 Caen c\'edex, France}
\altaffiliation{member of the Institut Universitaire de France}
   
\begin{abstract}
The presence of a phase transition in a finite system
can be deduced, together with its order, from the shape of the distribution
of the order parameter. This issue has been extensively studied in multifragmentation
experiments, with results that do not appear fully consistent. In this paper
we discuss the effect of the statistical ensemble or sorting conditions
on the shape of fragment distributions, and propose a new method, which can be easily implemented experimentally, to discriminate
between different fragmentation scenarii. This method, based on a reweighting of the measured distribution to account for the experimental constraints linked to the energy deposit, is tested on different simple models, and appears to provide 
a powerful discrimination.
\end{abstract}

\maketitle

\section{Introduction}
At the transition point of a first order phase transition, the distribution
of the order parameter in the corresponding finite system presents a characteristic
bimodal behavior in the canonical or grancanonical ensemble\cite{binder,topology,zeros,noi}.
The bimodality comes from an anomalous convexity of the underlying microcanonical entropy. It physically corresponds to the simultaneous presence of two different classes of events, which at the thermodynamic limit leads to phase coexistence.
Such behavior is very different from the signal of a second order phase transition, or more generally a critical behavior. In this latter case the order parameter distribution is generally wide but always monomodal, it fulfills specific scaling properties with increasing system size, and presents non-gaussian tails\cite{botet}.

In the case of nuclear multi-fragmentation, the most natural observable 
to analyze as a potential order parameter is the size of the heaviest cluster
produced in each collision event. 
Indeed this observable is known to provide an order parameter for a large class of transitions or critical phenomena involving complex clusters, from percolation to gelation, from reversible to irreversible aggregation\cite{botet}. 
Moreover it is reasonable to believe that the largest cluster size is always well correlated to the total energy deposit; this means that it will take different values in the two phases of any transition involving a finite latent heat, thus serving as an order parameter.
Many different indications exist\cite{wci2,moretto} that multi-fragmentation may be the finite size and possibly out-of-equilibrium realization of the liquid-gas
phase transition of infinite nuclear matter. Also in this case the heaviest 
cluster size is expected to play the role of an order parameter, because of its
good correlation with the system density, as shown by the numerical simulation
of the liquid-gas transition\cite{noi}.
 
These ideas have been recently applied by the INDRA collaboration to different
intermediate-energy heavy-ion collisions data sets, with the aim of tracking the multi-fragmentation phase transition and possibly determining its order\cite{lopez,pichon,borderie,bonnet,lautesse,frankland_prl,frankland}.
A systematic analysis of different data from central collisions\cite{frankland}
reveals that the heaviest fragment distributions are never bimodal, neither 
they present the non-gaussian tails that would allow to identify a critical phenomenon. The situation is different in peripheral collisions\cite{pichon}.
In this case, the quasi-projectile heaviest residues are sorted in centrality bins measured from the transverse energy of
light charged particles detected on the quasi-target side. The resulting distributions clearly show two different event families when plotted
against the asymmetry between the two heaviest fragments $Z_{asy}=(Z_{max}-Z_{second})/(Z_{max}+Z_{second})$, and do not fulfill the scaling properties shown by central collisions. The widths of the two partial distributions are too large for a bimodality to appear in the projection over 
the $Z_{max}$ axis, but a clear bimodality is seen in the closely
correlated $Z_{asy}$ distribution.

The characteristics and order of the nuclear 
multi-fragmentation transition are thus not completely clear.

It is certainly possible that two very different fragment formation
mechanisms act in central and peripheral collisions\cite{nicolas}.
It is however also important to note that the sorting conditions are not the
same in the different collisions, and the shape of the distributions is 
obviously influenced by the constraints imposed by the sorting.

In this paper we analyze the effect of the sorting, 
and propose a new method to detect
a possible first order phase transition independent of the mechanism of energy deposit, and without the artificial assumption of a heat bath.
Indeed the dependence on the sorting is a manifestation of the statistical
mechanics concept of ensemble inequivalence, as we develop in detail below.
Working out the relations between the different ensembles will allow us 
to predict the transformation to apply to the bimodality signal, in order to account for the experimental constraints.

\section{Bimodalities and ensembles}

\begin{figure}[htbp]
\begin{center}
\resizebox{0.80\columnwidth}{!}{\includegraphics{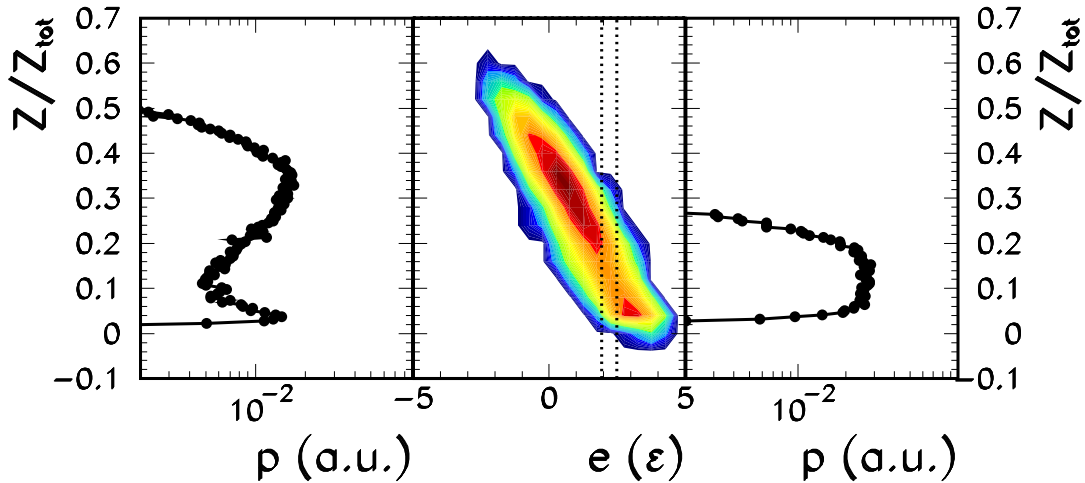}}
\end{center}
\caption{\it Event distributions in the isobar lattice gas model. 
A system of 216 particles is considered at a subcritical pressure 
and a temperature close to the transition temperature.
Central panel: canonical distribution in the largest 
cluster size and energy plane. The left and right figures
give the projection of this correlation over the largest cluster
size axis. Left: canonical ensemble; right: narrox constraint 
on the total energy as indicated by the two vertical lines of the
central panel.
Energy is given in units of the closest neighbors coupling.}
\label{fig:1}
\end{figure}

Observables distributions depend on the criterium used to collect the 
events, i.e. on the statistical ensemble.
Concerning the issue of bimodality,
a system experiencing a first order phase transition presents a bimodal
distribution of its order parameter if and only if the associated 
observable is constrained only in average through a conjugated Lagrange
parameter, and it is not fixed by a conservation law\cite{zeros,noi,carmona}.

For example, the energy distribution of a finite system, presenting 
in the thermodynamic limit a finite-latent-heat phase transition, is
bimodal in the canonical ensemble\cite{topology,labastie}. 
The same is true for any other observable $Z$ presenting a non-zero 
correlation with the total energy. If the correlation is loose and the 
system is small, each of the two peaks may have a large width, which may extend 
over values characteristic of the other phase, and mask the bimodality.
However, the distance between the two peaks $Z_1$ and $Z_2$ 
scales linearly with the number
of particles: $(Z_1-Z_2) \propto N$, while the width of each peak scales
only with its square-root: $\sigma_Z\propto \sqrt{N}$, meaning that a 
bimodal behavior will be recovered in the $Z$ observable, for sufficiently large systems. 
 
In the microcanonical ensemble, the energy distribution is by definition a $\delta$-function, and obviously cannot be bimodal. The distribution of any
other observable $Z$ will in general present a finite width, but if
the correlation with the energy is strong, the bimodality will be lost.
These general arguments are applied to the liquid-gas phase transition
in Fig.\ref{fig:1}, which shows the event distribution of the Lattice Gas Model
in the canonical isobar ensemble\cite{noi} at the transition temperature.
This model presents a transition with finite latent heat between a high density
liquid and a low density gas. The order parameter of the liquid-gas transition
is known too be one-dimensional, meaning that energy and density, or equivalently volume $V\propto\sum_i r_i^3$, have to be correlated.

This intuitive statement can be formalized in the framework of quantum
statistical mechanics\cite{reinhardt,wci_chomaz}.
Indeed the result of the statistical measurement of any observable $Z$,
associated to the operator $\hat{Z}$ is 

\begin{equation}
<\hat{Z}>_{\hat{D}}=\mathrm{Tr}\hat{Z}\hat{D},
\label{Eq:trAD}
\end{equation}

where $\hat{D}$ is the density matrix and $\mathrm{Tr}$ means the trace over the quantum Fock or Hilbert space. 
In the space of Hermitian matrices, 
the trace provides a scalar product \cite{reinhardt}

 \begin{equation}<<\hat{Z}||\hat{D}>>=\mathrm{Tr}\hat{Z}\hat{D}.  \label{EQ:scalarAD}
\end{equation}

It is then possible to define an orthonormal basis of Hermitian operators $\{ \hat{Z}_{l}\} $ in the observables space, and to interpret the measurement $<\hat{Z}_{l}>_{\hat{D}}$ as the projection of the density matrix 
over the corresponding axis. Statistically independent observations can thus 
be associated to orthogonal observables in the Liouville space.

As a practical consequence, in the liquid-gas phase transition the distribution
at the transition temperature must be bimodal both in the volume and in the energy 
direction\cite{noi}.  As anticipated in the introduction, the size of the largest cluster
is also correlated to both volume and energy, and as such it can be used as an equivalent order parameter. This is shown in the left side of Fig.\ref{fig:1}: 
at the transition temperature, the distribution of the largest-cluster size 
shows two peaks separated by a region of convexity. 
These observations are done in the canonical ensemble.
If energy is constrained as in the microcanonical ensemble, the two phases are 
not accessible in the same ensemble of events any more, and bimodality is lost.
This is not only true for the exact energy conservation implied by the 
microcanonical ensemble, but for any strong constraint on the energy.
This is shown in the right side of Fig.\ref{fig:1}: the distribution issued
from a narrow energy constraint on the canonical event set is a normal distribution.

\section{Bimodalities and convexity anomalies}

This behavior is a consequence of ensemble inequivalence, 
which has been addressed by numerous works in the literature\cite{inequivalence}.
In particular the presence of a conservation constraint on an order parameter
has been reported to modify different phase transition observables, making a first order transition look like a continuous one\cite{richert,carmona,big}. 

The prototype of an ensemble constraining the order parameter is the microcanonical ensemble. There, the phase transition can be univocally recognized
studying the curvature properties of the density of states\cite{gross}.
Indeed from the standard definition of the canonical ensemble
\begin{equation}
p_{\beta}(E) = W(E) \exp (-\beta E ) Z_{\beta}^{-1}
\label{distri_cano}
\end{equation}
we can immediately see that a bimodality in the canonical energy 
distribution is exactly equivalent to a convex intruder in the microcanonical
entropy $S=\log W$, which leads to the well known microcanonical 
negative heat capacity\cite{noi,gross}.
Let us now consider the case of a second observable $Z$.
If both $E$ and $Z$ are order parameters, and the transition is first order, the 
two-dimensional probability distribution $p(E,Z)$ should be bimodal
in both the $E$ and the $Z$ direction within the ensemble where the 
observables are fixed by the conjugated 
Lagrange multipliers $\beta,\lambda$ \cite{zeros}:
\begin{equation}
p_{\beta}(E,Z) = W(E,Z) \exp (-\beta E - \lambda Z ) Z_{\beta,\lambda}^{-1}
\label{distri_cano_2D}
\end{equation}
All conservation laws on other variables are implicitly accounted
in the definition of the density of states $W$. For instance, if $Z$ represents
the largest cluster size and the total system size is $Z_{tot}$,
$W$ reads
\begin{equation}
W(E,Z)=W_{Z_{tot}}(E,Z)=\sum_{(n)} \delta(E-E^{(n)}) \delta(Z-Z^{(n)}) \delta(Z_{tot}-Z_{tot}^{(n)})
\end{equation}
where the sum extends over the system microstates.
The search for bimodalities can only be done in this (extended)
canonical ensemble, and is exactly equivalent to the study 
of the curvature matrix of the entropy in the two-dimensional
observable space
\begin{equation}
C=
\left( 
\begin{array}{ll}
\partial^2 S/ \partial E^2 & \partial^2 S/ \partial E\partial Z \\ 
\partial^2 S/ \partial Z \partial E & \partial^2 S/ \partial Z^2\\ 
\end{array}
\right)
\label{curvature}
\end{equation}
If this curvature matrix has two positive eigen-values,
this means that and $Z$ and $E$ are associated to two independent order parameters.
In the more physical case of a one-dimensional order parameter,
only one eigen-value is positive, and the associated eigen-vector
can be taken as the "`best"' order parameter. It is the linear combination
of the $E$ and $Z$ observables, which gives the best separation 
of the two phases in the two-dimensional space\cite{topology}.

In the physical case of nuclear multi-fragmentation experimentally 
studied through nuclear collisions, the distribution of the deposited energy crucially depends on the entrance channel dynamics and data selection criteria. In the case of quasi-projectile events selected in heavy ion collisions,
the energy distribution is very large, and is determined by the impact parameter geometry and dissipation dynamics. If events are sorted in centrality 
bins, the distribution is centered on a well defined value given by 
the average dissipation at the considered impact parameter, but the 
distribution has a finite width that depends in a non-controlled way
on the selection criteria. The statistical ensemble describing 
multi-fragmentation data is thus neither canonical nor microcanonical,
and should rather be described in terms of the gaussian ensemble\cite{challa},
which gives a continuous interpolation between canonical and microcanonical.
If $E$ and $Z$ have a non zero correlation as we expect
, $W(E,Z)\neq W_E(E) W_Z(Z)$,
the distribution of energy will affect also the distribution of $Z$,
and the concavity of the $Z$ distribution will not be univocally linked to the concavity 
of the entropy. 

Concerning the explicit constraint $\lambda$ on $Z$, 
there is no reason to believe that the collision dynamics or the data treatment 
induces a specific constraint on the size of the largest cluster, 
other than the total mass and charge conservation, which are
already implemented in the definition of the state density. This means
that we will consider in the following $\lambda=0$. 
  
\section{Relying observables distributions to the underlying entropy}

In the last section we have argued that, in the absence of a canonical sorting, there is no one-to-one correspondence between the shape of the probability distribution and the phase transition properties of the system implied by its 
density of states. This is not only true for the energy, but also 
for all other observables that present a non-zero correlation with the energy.
Unfortunately, if the order parameter is one-dimensional, all  
observables that can be proposed to study bimodality (charge of the heaviest cluster, asymmetry, etc..) must be correlated with the energy. 
This means that the presence (or absence) of the 
bimodality signal may depend on the experimental sorting conditions.

It is however important to note that if the energy distribution 
cannot be experimentally controlled, it can be - at least approximately -
a-posteriori measured. This means that it is possible to unfold 
from the experimental distribution the contribution of the entropy, giving 
the phase properties of the system, and 
the contribution of the energy distribution, which depends on the collision
dynamics.
Indeed, as long as no explicit bias acts on the $Z$ variable, 
the experimental distribution can be calculated from the canonical one 
eq.(\ref{distri_cano_2D}) by a simple reweighting of the probabilities associated to each deposited energy
\begin{equation}
p_{exp}(E,Z)=p_{\beta}(E,Z)\frac{p_{exp}(E)}{p_{\beta}(E)}
\label{gexp}
\end{equation}
where $p_{exp}(E)$ is the measured energy distribution.

In the following, we explicit the effect of this experimental bias 
in the two cases of interest: an entropy  curvature matrix with two negative eigen-values, corresponding to the absence of a phase transition, and an entropy presenting one direction of convexity (first order phase transition).

\subsection{The monomodal case}

\begin{figure}[htbp]
\begin{center}
\resizebox{0.60\columnwidth}{!}{\includegraphics{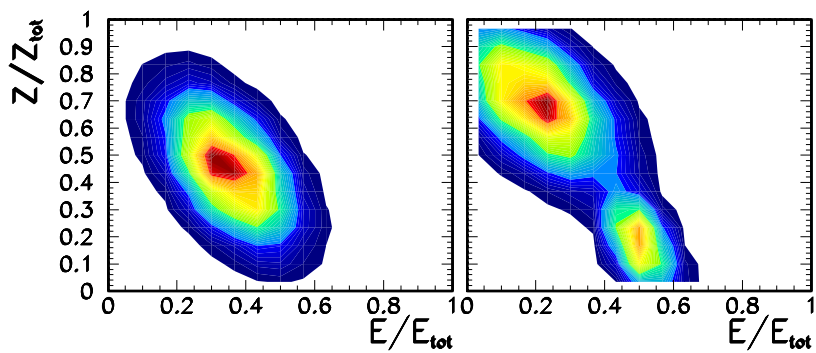}}
\end{center}
\caption{\it Canonical distributions in the plane of the total thermal
energy, normalized to the total available energy, and the size of the largest cluster, normalized to the total system size, in the (double) saddle point approximation.
Right part: case of a first order phase transition eq.(\ref{gcano_t}).
Left part: without transition, eq.(\ref{gcano}). 
The corresponding parameters
are given in the text. }
\label{fig:2}
\end{figure}

Let us consider again the density of states $W(E,Z)=\exp(S(E,Z))$ 
in the two-dimensional observable
space defined by the total energy $E$ and the largest cluster size $Z$.
In the absence of a phase transition, this function is concave everywhere.
To evaluate the partition sum at a temperature $\beta^{-1}$ 
we can make a saddle point approximation
\begin{equation}
S(E,Z)=S_0+\beta(E-E_{\beta})-c_{11}(E-E_{\beta})^2
-c_{22}(Z-Z_{\beta})^2+c_{12}(E-E_{\beta})(Z-Z_{\beta})
\label{saddle}
\end{equation}
where $E_{\beta}$,$Z_{\beta}$ are the most probable values of $E$ and $Z$
at a temperature $\beta^{-1}$, and 
$C=\{c_{ij}\}$ is the entropy curvature matrix eq.(\ref{curvature}).
In this approximation the event distribution is a gaussian
\begin{equation}
p_{\beta}(E,Z)=\frac{1}{2\pi}\frac{1}{\sqrt{det\Sigma}}
\exp\left ( -\frac{1}{2}\vec{x}\Sigma^{-1}\vec{x}\right)
\label{gcano}
\end{equation}
where $\vec{x}=\left (E-E_{\beta},Z-Z_{\beta}\right )$, 
and the variance-covariance matrix $\Sigma$
is related to the curvature matrix by 
\begin{eqnarray}
\frac{1}{2(1-\rho^2)\sigma_E^2}&=&c_{11}=-\frac{1}{2}
\frac{\partial^2S}{\partial E^2}|_{E_{\beta}} \\
\frac{1}{2(1-\rho^2)\sigma_Z^2}&=&c_{22}=-\frac{1}{2}
\frac{\partial^2S}{\partial Z^2}|_{Z_{\beta}} \\
\frac{\rho}{(1-\rho^2)\sigma_E\sigma_Z}&=&c_{12}=c_{21}=\frac{1}{2}
\frac{\partial^2S}{\partial E \partial Z}|_{E_{\beta},Z_{\beta}} 
\end{eqnarray}
An example is given in the left part of Figure \ref{fig:2}.
To fix the ideas we have chosen $E_{\beta}=0.35 E_{tot}$,
$Z_{\beta}=0.45 Z_{tot}$, $\sigma_E=0.1 E_{tot}$, 
$\sigma_Z=0.15 Z_{tot}$, $\rho=-0.6$, where $E_{tot}$ and $Z_{tot}$ are the 
total available energy and size, respectively.
The inclusive probabilities associated to this distribution, or marginal
distributions, are simple gaussians
\begin{eqnarray}
p_{\beta}(E)&=&\frac{1}{\sqrt{2\pi \sigma_E^2}} 
\exp\left ( -\frac{(E-E_{\beta})^2}{2 \sigma_E^2}\right) 
\label{gcano_e} \\
p_{\beta}(Z)&=&\frac{1}{\sqrt{2\pi \sigma_Z^2}} 
\exp\left ( -\frac{(Z-Z_{\beta})^2}{2 \sigma_Z^2}\right) 
\label{gcano_z}\\
\end{eqnarray}
Deviations from a perfect
gaussian are in principle possible because of the effect of the boundary,
which limits the possible values of $E$ and $Z$ due to the conservation laws.
These effects are however very small in all physical cases of interest, and 
the marginal distributions are largely independent of the correlation coefficient.

\begin{figure}[htbp]
\begin{center}
\resizebox{0.70\columnwidth}{!}{\includegraphics{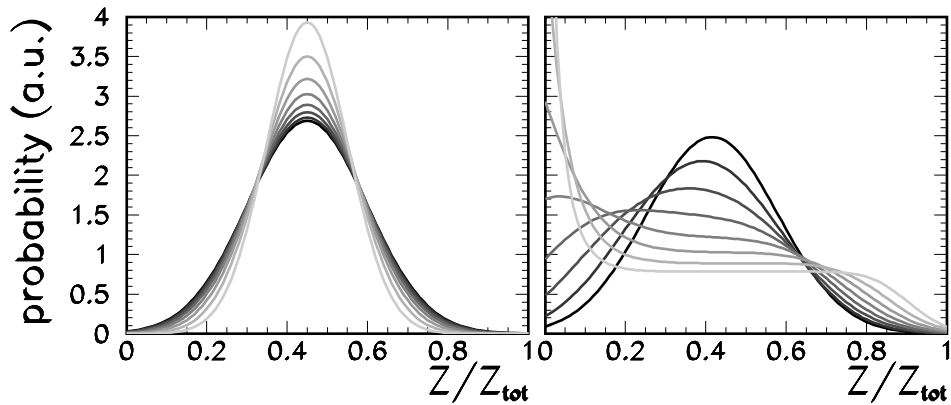}}
 \end{center}
\caption{\it 
Marginal distribution for the monomodal case
represented in the left part of Figure \protect\ref{fig:2}. 
Left: gaussian ensemble eq.(\protect\ref{gexp}). 
Right: reweighted distribution eq.(\protect\ref{pw}). 
The correlation coefficient is varied 
from $\rho=-0.15$ (black lines) to $\rho=-0.85$ (lightest grey).
The other parameters are taken as for Figure \protect\ref{fig:2}.
}
\label{fig:3}
\end{figure}

As we have stressed in the previous section,
in the experimental case, because of the absence of an external heat
bath, there is no reason to believe that the event distribution can 
be described by eq.(\ref{gcano}). Even if the partitions may fully explore
the available phase space, the dynamics of the
entrance channel plus the detection and sorting constraints will lead
to an energy distribution which is not uniquely characterized by its 
average value $<E>$. When data are sorted in centrality bins, 
the energy distributions as measured from calorimetric algorithms
are generally close to gaussians. 
This can be easily understood in the framework of information theory\cite{bal82}, 
where such a situation can be dealt with introducing 
the energy variance as a second constraint,
which naturally leads to the gaussian ensemble\cite{challa}.
Therefore we will take as a typical shape for 
$p_{exp}(E,Z)$ the form (\ref{gexp}) 
where $p_{exp}(E)$ is a gaussian of average $E_{exp}$ and variance 
$\sigma^2_{exp}$. This choice is only done to present some realistic
numerical applications, and is readily extended to more general
situations.

Eq.(\ref{gcano}) and eq.(\ref{gexp}) have 
the same functional dependence, with an important difference though.  
In the case of eq.(\ref{gcano}) the first two
moments of the energy distribution eq.(\ref{gcano_e}) are directly related to 
the canonical partition sum $\ln Z_\beta$ by
\begin{eqnarray}
E_{\beta}&=&-\frac{\partial \ln Z_{\beta}}{\partial \beta} \\
C \beta^{-2} &=& \sigma^2_{E}=\frac{\partial^2 \ln Z_{\beta}}{\partial \beta^2} 
\end{eqnarray}
and, in the saddle point approximation, to the microcanonical temperature $T$
and heat capacity $C_\mu$
\begin{eqnarray}
T^{-1}&=&\frac{\partial S(E,Z)}{\partial E}|_{E_{\beta}} \\
\frac{1}{(1-\rho^2)\sigma^2_{E}}&=&
\frac{1}{C_\mu T^2}= -\frac{\partial^2 S(E,Z)}{\partial E^2}|_{E_{\beta}} .
\end{eqnarray}
On the other hand, the first two
moments of the energy distribution (\ref{gexp}) are external constraints
imposed by the collision dynamics, that have a priori no direct 
connection with thermodynamics.
Since the folding of gaussians is still a gaussian,
the corresponding $Z$ distributions, shown for the model case
discussed above  in the left part of Figure \ref{fig:3},
are also of gaussian shape.
The width of the distribution depends on the value of the correlation
coefficient $\rho$ (a lower value of the correlation leading to a larger
distribution), and all distributions are normal. For the specific application 
of Figure \ref{fig:3}, we have fixed $E_{exp}=E_{\beta}$, 
$\sigma_{exp}=\sigma_E/2$.

These observations mean that, if the system does not present a phase transition, the non-canonical experimental sorting does
not qualitatively bias the shape of the measured $Z$ distribution.

\subsection{The bimodal case}

\begin{figure}[htbp]
\begin{center}
\resizebox{0.70\columnwidth}{!}{\includegraphics{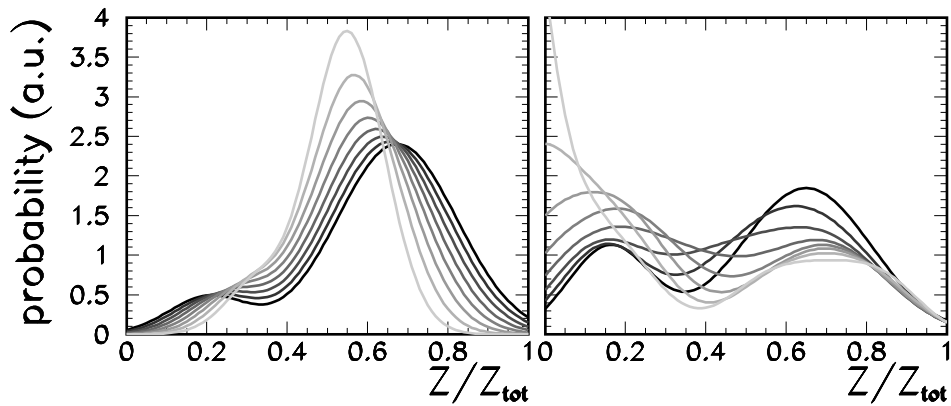}}
\end{center}
\caption{\it 
Marginal distribution for the bimodal case
represented in the right part of Figure \protect\ref{fig:2}. 
Left: gaussian ensemble eq.(\protect\ref{gexp_z}). 
Right: reweighted distribution eq.(\protect\ref{pw}).
The correlation coefficient is varied 
from $\rho=-0.15$ (black lines) to $\rho=-0.85$ (lightest grey). 
The other parameters are taken as for Figure \protect\ref{fig:2}.
}
\label{fig:3bis}
\end{figure}
 
In the presence of a first order phase transition with a finite 
latent heat, the saddle point 
approximation eq.(\ref{saddle}) fails\cite{gross}. 
If the system is not too small,
a double saddle point approximation around the two maxima 
$(E_l,Z_l)$, $(E_g,Z_g)$ can however 
be employed \cite{zeros}, leading to a double humped distribution
\begin{equation}
p_{\beta}(E,Z)= N_l \frac{1}{\sqrt{det\Sigma_l}}
\exp\left ( -\frac{1}{2}\vec{x_l}\Sigma_l^{-1}\vec{x_l}\right)
+ N_g \frac{1}{\sqrt{det\Sigma_g}}
\exp\left ( -\frac{1}{2}\vec{x_g}\Sigma_g^{-1}\vec{x_g}\right)
\label{gcano_t}
\end{equation}
where $\vec{x_i}=\left (E-E_i,Z-Z_i\right )$,$i=l,g$, $\Sigma_l$ 
($\Sigma_g$) represent the variance-covariance matrix evaluated 
at the liquid (gas) solution, and $N_l(\beta)$,$N_g(\beta)$ 
are the proportions of the two phases
, with $N_l\sqrt{det\Sigma_g}=N_g\sqrt{det\Sigma_l}$ at the transition temperature $\beta_t$.
An example of a distribution at $\beta_t$ 
is given in the right part of Figure \ref{fig:2}.
To fix the ideas we have chosen $E_{l}=0.2 E_{tot}$,
$Z_{l}=0.7 Z_{tot}$, $\sigma_{El}=0.12 E_{tot}$, 
$\sigma_{Zl}=0.15 Z_{tot}$,$E_{g}=0.5 E_{tot}$,
$Z_{g}=0.18 Z_{tot}$, $\sigma_{Eg}=0.05 E_{tot}$, 
$\sigma_{Zg}=0.1 Z_{tot}$, $\rho_l=\rho_g$. 

If the distance in the $Z$ direction between the two phases $Z_l-Z_g$ 
is sufficiently large respect to the associated variances $\sigma_{Zl}$, 
$\sigma_{Zg}$, as it is the case in Fig.\ref{fig:2}, 
the presence of the transition can then be inferred
looking for the temperature interval associated to a bimodal 
shape of the $Z$ distribution\cite{topology}
\begin{equation}
p_{\beta}(Z)= N_l \sqrt{\frac{2\pi}{\sigma^2_{Zl}}}
\exp\left ( -\frac{(Z-Z_l)^2}{2\sigma^2_{Zl}} \right)
+ N_g \sqrt{\frac{2\pi}{\sigma^2_{Zg}}}
\exp\left ( -\frac{(Z-Z_g)^2}{2\sigma^2_{Zg}} \right) 
\label{gcano_zt}
\end{equation}
%

If we do not dispose of a canonical sample however the situation 
is not so simple. Let us consider the case where
the experimental energy distribution belongs to the gaussian ensemble
with an average $E_{exp}$ and a variance $\sigma^2_{exp}$ which 
can be experimentally measured for each given sorting.
Then the measured $Z$ distribution reads (taking $\rho_l=\rho_g$):
\begin{equation}
p_{exp}(Z) \propto \int_{E_{min}}^{E_{max}} dE \;
p_{\beta_t}(E,Z)\frac
{\exp\left ( -\frac{(E-E_{exp})^2}{2 \sigma_{exp}^2}\right) }
{\frac{\sigma_{Zl}}{  \sigma_{El}\sigma_{Zl}+
\sigma_{Eg}\sigma_{Zg}}  
\exp\left ( -\frac{(E-E_l)^2}{2\sigma^2_{El}} \right)
+ \frac{\sigma_{Zg}}{  \sigma_{El}\sigma_{Zl}
+\sigma_{Eg}\sigma_{Zg}}  
\exp\left ( -\frac{(E-E_g)^2}{2\sigma^2_{Eg}} \right) }
\label{gexp_z}
\end{equation}
which may look close to  gaussian even in the phase transition
region. In particular if the energy distribution is not very 
large,  the sample corresponding
to the transition energy $E_t$ (i.e. the energy corresponding to the 
minimum probability of eq.(\ref{gcano_t})) will not explore
the energy domains corresponding to the liquid 
$E \approx E_l \pm \sigma_{El}$ 
and to the gas phase $E\approx E_g \pm \sigma_{Eg}$.   
If $Z$ is well correlated to $E$ as we expect, then the only size 
partitions
explored at the transition energy will be the intermediate  
ones between the liquid and the gas, leading to a normal $Z$ distribution.
This can be clearly seen in the left part of Fig.\ref{fig:3bis}. 
Only for very low values of  the correlation coefficient 
the presence of the two phases can be recognized from the 
$Z$ distribution, where the minimal value for $\rho$ depends on the width
of the inclusive energy distribution in the specific experimental situation under study.

\section{Reweighting the energy distribution}

In order to recover the information on the concavity of the entropy
from the measured distribution eq.(\ref{gexp_z}), we have to get rid
of the dominant energy dependence $p_{exp}(E)$.
To this aim we can reweight the total $(E,Z)$ distribution in such a 
way that the total energy distribution is a constant between $E_{min}$
and $E_{max}$, with $E_{min}<E<E_{max}$, and $E_{min}$, $E_{max}$ 
are chosen large enough, such that the spanned energy interval contains the two phases. 
This procedure corresponds
to introducing a weight for each $E$ bin
\begin{equation}
w(E)=\left ( \int_0^{Z_{max}} dZ \; p_{exp}(E,Z) \right )^{-1}
\end{equation}
and generates a new weighted distribution which is a direct measure 
of the density of states:
\begin{equation}
p_w(E,Z) = \frac{p_{\beta_t}(E,Z)}{p_{\beta_t}(E)}=
\frac{W(E,Z)}{W(E)}.
\label{pw}
\end{equation}

In the limiting case of a perfect (anti)correlation 
between $E$ and $Z$, the concavity anomaly in the two-dimensional 
entropy $S(E,Z)$ is solely due to the convex intruder in the function
$S(E)$. In this case the order parameter is aligned with the $E$
direction, and it is clear that the reweighted distribution eq.(\ref{pw})
will be a normal distribution in spite of the presence of a first
order phase transition. However, in the case of the liquid 
gas phase transition we have good reasons to expect\cite{europhys00}
the order parameter in the enthalpy $H=E+pV$ direction. 
Since $Z$ is correlated to the volume as well as to the energy, then
we may expect a non zero component of the order parameter 
along the $Z$ axis, and an imperfect correlation between $Z$ and $E$.
In this case an anomaly in $S(E,Z)$ may persist even after 
subtraction of the entropy $S(E)$, and this anomaly will be shown
by a residual bimodality in the distribution (\ref{pw}).

\subsection{Application to model cases}

The projections of $p_w$ in the $Z$ direction are shown
in a perfectly monomodal and perfectly bimodal case, in the right parts of Figs.\ref{fig:3} and \ref{fig:3bis} respectively.
If the underlying entropy presents a convex intruder (Fig.\ref{fig:3bis}),
the bimodality persits in the reweighted distribution
for all considered values of the 
correlation coefficient.
The case of a normal distribution is explored in Fig.\ref{fig:3}: 
the distribution is still approximately gaussian 
if the correlation coefficient is small,
while if $\rho$ is close to $-1$ it tends to a constant,
but in no case the reweighting procedure applied to the 
monomodal distribution creates a spurious bimodality. 

\begin{figure}[htbp]
\begin{center}
\resizebox{0.80\columnwidth}{!}{\includegraphics{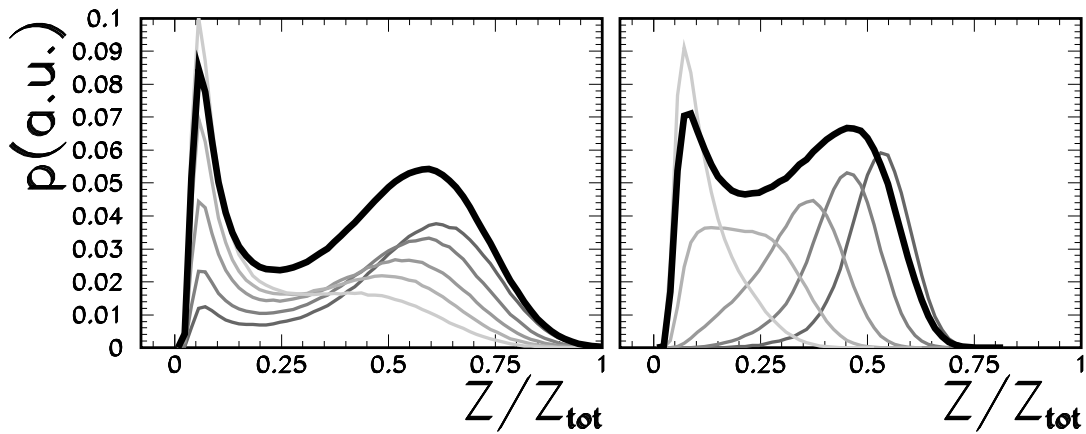}}
 \end{center}
\caption{\it Distributions of the size of the largest
cluster for a system of N=216 particles within 
the isobar lattice gas model at subcritical pressure.
Left: canonical calculations at temperatures $T=3.72,3.74,3.76,3.78,3.80$ MeV
close to the transition temperature (thin lines), and 
their sum (thick line) obtained weighting evenly the different
temperatures. Right: microcanonical distributions at energies spanning the coexistence region $E/N=0.75,1.25,1.75,2.25,2.75$ MeV
(thin lines), and their sum (thick line) obtained weighting evenly the different
energies. The closest neighbors coupling is fixed to $\epsilon=5.5$ MeV.
}
\label{fig:4}
\end{figure}

Another example of the effect of the reweighting procedure 
is given in Fig.\ref{fig:4}.
The right part of Fig.\ref{fig:4} shows microcanonical calculations
within the lattice gas model\cite{noi}. Several distributions of the largest
cluster size obtained for different total energies are displayed.
In agreement with Fig.\ref{fig:1}, these distributions are never bimodal,
even if the total energy interval is chosen such that the first order 
phase transition is crossed by the calculation. The bimodality is recovered
if these distributions are summed up with equal weight following eq.(\ref{pw})
(thick line in Fig.\ref{fig:4}).
The detailed shape of the reweighted distribution obviously depends on 
the energy interval used, a wider interval leading to an increased distance in energy between the maximum and the minimum.

For comparison, canonical distributions at different temperatures close
to the transition temperature are displayed in the left part of the same figure.
Because of the finite system size, the distributions are bimodal over a 
narrow but finite temperature range, while the transition temperature
can be identified from the requirement that the two peaks have approximately 
the same height\cite{noi}. We can see that the canonical distribution close to 
the transition point is wider than the reweighted microcanonical one, and its concavity anomaly is more pronounced, however the shapes of the two distributions are very close.
The width of the transition interval (i.e. the distance between the two peaks) 
as estimated from microcanonical calculations is a lower limit of the physical transition interval, and the quality of this estimation depends on the energy width of the available sample.

In the experimental case, due to the incomplete detection and the imperfect 
emission source selection, only an approximation
of the deposited energy can be measured via measurable observables like transverse
energy or calorimetric energy.
We have checked that allowing a finite energy width does not change the result:
an equal reweighting of gaussian ensembles leads to a distribution virtually identical to the one displayed in the right panel of Fig.\ref{fig:4}. An equal reweighting of canonical distributions is also shown in Fig.\ref{fig:4}: the resulting distribution is almost superposable to the distribution at the transition temperature.

\begin{figure}[htbp]
\vskip -1.cm
\begin{center}
\resizebox{0.50\columnwidth}{!}{\includegraphics{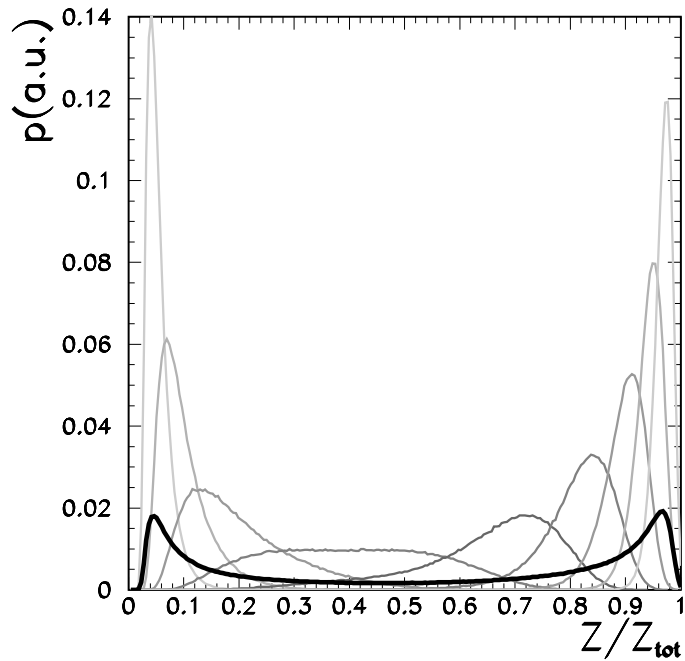}}
 \end{center}
\caption{\it 
Calculations in the bond three-dimensional percolation model
for a system of 216 particles. 
Thin lines: distribution of the size of the largest 
cluster for equally spaced values of the bond-breaking probability 
from $p=.16$ to $p=.56$ including the critical value $p_c$.
Thick line: summed distribution obtained weighting evenly the different
$p$'s.
}
\label{fig:5}
\end{figure}

These examples suggest that, if the microcanonical entropy presents a convexity,
it may be possible to recognize it from the experimental data through a reweighting procedure,
provided we dispose of an experimental sample sufficiently large in deposited
energy. 

A more challenging situation is given by the presence of a critical phenomenon.
In this case a variation of the control parameter leads again the system to pass from one phase to another, but this transition is continuous and does not correspond to a convex microcanonical entropy.
Fig.\ref{fig:5} shows calculations performed within the bond-percolation
model\cite{stauffer}. This model presents 
a critical behavior at a value $p_c$ of the 
bond-breaking probability. As the Lattice gas model can be considered as a prototype for phase coexistence and first-order phase transitions, similarly
the percolation model is the simplest realization of a second order phase transition. 
The percolation distribution close to the critical point 
is very close to the microcanonical lattice gas distribution in the
middle of the coexistence region of the first-order transition.
The consequence of this well-known feature\cite{richert,big} is that 
it is still not clear whether nuclear-multifragmentation 
is better described as a critical phenomenon\cite{campi} or 
a first-order phase transition\cite{michela,bruno}.

Figure \ref{fig:5} shows that also in this case 
the presence of a phase transition is detected by the reweighting 
procedure: a flat distribution of bond-breaking probabilities 
leads to a global distribution 
qualitatively very similar to the distribution at the critical point
if the p's interval is narrow, but if we have a sample including 
p's sufficiently far away from the critical probability as in Figure \ref{fig:5}, two peaks appear
corresponding to the ordered ($p\approx 0$) and desordered ($p\approx 1$)
phase (thick line).

\begin{figure}[htbp]
\begin{center}
\resizebox{0.80\columnwidth}{!}{\includegraphics{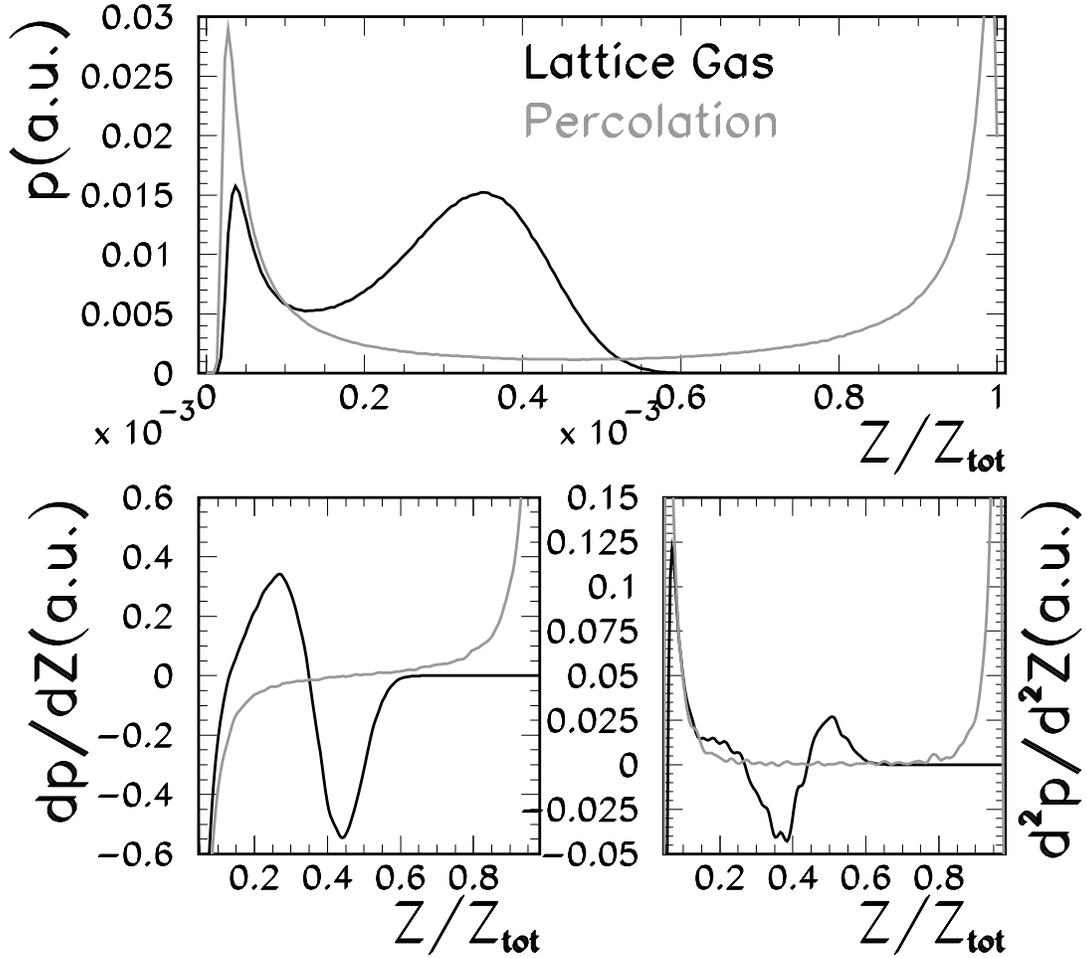}}
 \end{center}
\caption{\it Upper part: reweighted distributions of the size of the largest
cluster for a system of N=216 particles with the lattice gas and the percolation
model.
Lower part: first (left) and second (right) derivatives of the distributions.
}
\label{fig:6}
\end{figure}

From Figure \ref{fig:4} and Figure \ref{fig:5}
we can see that in both models the reweighted distribution 
show the presence of the two phases, with an important difference though.
If the lattice-gas calculation
shows a clear convexity between the two phases, associated to the convex intruder
in the microcanonical entropy, this is not the case for the percolation distribution which, in the critical mass region, largely keeps the convexity  
of the critical point, i.e. it is essentially flat.
This statement is quantified in Figure \ref{fig:6}, which shows the numerically
calculated first and second derivative of the reweighted distributions in the 
two models. Only the model associated to a first order phase transition 
exhibits a region of positive convexity.

This result can be easily understood. Indeed from eq.(\ref{pw})
we can identify the (inverse) curvature of the reweighted distribution with a 
(generalized) susceptibility:
\begin{equation}
\frac{\partial^2 \log p_w}{\partial Z^2} = 
\frac{\partial^2 S}{\partial Z^2} = \chi_{\mu}^{-1}
\label{susceptibility}
\end{equation}

The back-bending shown by the lattice gas calculation in Fig.\ref{fig:6}
has then the same physical meaning as the well-known back-bending
of the microcanonical caloric curve\cite{gross}, and allows to unambiguously
identify a discontinuous (first order) phase transition.

\subsection{Application to experimental data}

In the previous section we have seen that, for different model cases 
that can be relevant for nuclear multi-fragmentation, 
the concavity properties of the reweighted distribution 
reflect the concavity of the underlying entropy.

However we may wonder whether the different behaviors 
shown in Figs.\ref{fig:3},\ref{fig:3bis},\ref{fig:4},\ref{fig:5} will be distinguishable in realistic
experimental cases, where distributions are often affected
by important statistical as well as systematic error bars.

Even more important, we may also ask whether the behaviors 
of the schematic models we have shown can be taken as general
representatives of the presence or absence of a first order
phase transition.
From a mathematical point of view, the two-dimensional function 
$\log p_w(x_1,x_2)$ is concave 
everywhere if and only if all the eigenvalues
of the curvature matrix 
$c'_{ij}=\partial^2\left (S(x_1,x_2)-S(x_1)\right )/\partial x_i
\partial x_j$, are negative. A necessary and sufficient condition 
that an hermitian matrix has negative eigenvalues is that the minors
of the determinant are negative, $i.e.$
\[
c'_{11}\leq 0 \;\;\; ; \;\;\; detC'\leq 0
\]
which in our case becomes
\begin{equation}
\frac{\partial^2 S(E,Z)}{\partial E^2}-\frac{\partial^2 S(E)}{\partial E^2}\leq
0 \;\; ; \;\;  detC - \frac{\partial^2 S(E)}{\partial E^2}
\frac{\partial^2 S(E,Z)}{\partial Z^2} \leq 0
\end{equation}
for every $(E,Z)$ value.
In principle these conditions can be 
violated even if $S(E,Z)$ and $S(E)$ are concave everywhere.
This means that we cannot a priori exclude a concave entropy even
if we observe a bimodality in $p_w$.  

In addition to that, only if a distribution is approximately 
symmetric we can visually judge, or numerically calculate, its convexity properties. In the experimental case, we may not 
dispose of an energy sample wide enough to construct a symmetric distribution around the transition value. It is clear for exemple that
if in Figs.\ref{fig:4}, \ref{fig:5} we would sum up only 
low energy (respectively, low $p$s) samples, 
in both models a shoulder at low $Z$ 
would be observed, and we would not be able to clearly discriminate
between the first order and the second order scenario. 
This is probably the case in quasi-projectiles events,
where the dynamics of the collision strongly favors low excitation energy 
deposit, and the desordered phase may barely, if ever, be attained\cite{pichon}.

To be conclusive about the presence 
of a first order phase transition we have therefore 
to show that the observed distribution is not compatible
with a concave entropy, or in other words that a single
saddle point approximation cannot explain the data 
as we have developed in section 4.

In realistic experimental cases this can be tested through a $\chi^2$ test.
Indeed in the absence of a phase transition the underlying entropy is everywhere
concave, a single saddle point approximation is reasonable, and the 
reweighted two-dimensional probability should be well described by:

\begin{equation}
p_w(E,Z)= \frac{W(E,Z)}{W(E)}=\frac{1}{\sqrt{2\pi\sigma_Z^2(1-\rho^2)}}
exp\left [ - \frac{1}{2(1-\rho^2)} \left ( \frac{\rho}{\sigma_E}(E-E_0)
- \frac{1}{\sigma_Z}(Z-Z_0)
\right )^2 \right ] \label{fit}
\end{equation}

The parameters $E_0,Z_0,\sigma_E,\sigma_Z,\rho$ should be fitted 
to the experimental reweighted distribution. A $\chi^2$ test should allow
to exclude the adequacy of the fit.

If fragmentation events are compatible with a critical phenomenon,
both the single and the double saddle point approximation should fail.
This means that in this scenario we should not be able to describe the data
with the reweighted single gaussian eq.(\ref{fit}), not with the 
reweighted double gaussian.
An application of this method to INDRA data is in progress\cite{bonnet}.

\section{Conclusions}

In this paper we have critically analyzed the bimodality observable
proposed in \cite{topology} to track a first order phase transition,
which has been extensively used to analyze multi-fragmentation data\cite{lopez,pichon,borderie,bonnet,lautesse}.
We have shown that the out-of-equilibrium dynamics of the entrance
channel, as well as the different data sorting conditions, can bias 
the signal in an uncontrolled way, tending to suppress the bimodal 
behavior. Based on the correlation between the bimodality observable
and the deposited energy, we propose a reweighting method on the measured
distributions, with the aim of disentangling the dynamic dissipation 
properties from the entropy characteristics. This method applied on 
different schematic models, is seen to provide an excellent discrimination
between the case of a first-order phase transition, a critical phenomenon
and the absence of transition.
A chi-square test is proposed for the experimental case to demonstrate/refute 
the existence of a convexity in the microcanonical entropy.

\end{document}